# Experimental and Numerical Validation of Tape-Based Metasurfaces in Guiding High-Frequency Surface Waves for Efficient Power Transfer

K. Suzuki[1], P. T. Dang[1], H. Homma[1], A. A. Fathnan[1], Y. Ashikaga[2], Y. Tsuchiya[2], S. Phang[3], and H. Wakatsuchi[1]

[1]*Department of Engineering, Nagoya Institute of Technology, Gokiso-cho, Showa, Nagoya, Aichi, 466-8555, Japan*

[2]*Teraoka Seisakusyo Co., Ltd., 1-4-22 Hiromachi, Shinagawa-ward, Tokyo, 140-8711, Japan*

[3]*George Green Institute for Electromagnetics Research, Faculty of Engineering, University of Nottingham, University Park, Nottingham NG7 2RD, UK*

We present an effective method for transmitting electromagnetic waves as surface waves with a tape-based metasurface design. This design incorporates silver square patches periodically patterned on an adhesive tape substrate. Specifically, our study proposes a strategy to enhance the efficiency of power transfer in high-frequency bands by guiding signals as surface waves rather than free-space waves. Both the numerical and experimental results validate the markedly enhanced efficiency in power transfer of high-frequency signals compared to that achieved with conventional methods, such as wireless power transfer and microstrips. Importantly, our metasurface design can be readily manufactured and tailored for various environments. Thus, our study contributes to designing power-efficient next-generation communication systems such as 6G and 7G, which leverage high-frequency signals in the millimeter-wave and THz bands.

Metasurfaces (MSs) are two-dimensional or planar versions of metamaterials that feature a subwavelength-scale thickness and unit cell.[1-5] These structures have attracted widespread attention within the field of electromagnetics thanks to their lightweight properties and simplified manufacturing processes. Their distinct capabilities encompass the manipulation of waves, including blocking, absorption, concentration, dispersion, or guiding across surface planes of waves at incident angles and within space of waves at normal and oblique angles for frequencies spanning from microwaves to those within the visible spectrum.[6-9] MSs operate on a fundamental principle involving arrays of periodically placed resonators with subwavelength lengths and varied geometric parameters (shape, size, and orientation). These configurations create diverse optimal responses by engineering wave propagation properties such as the phase,[10] amplitude, polarization,[11] and impedance[12] to actively control wave propagation.[4,13] Although past research mainly focused on physical aspects[14,15] and on applications of MSs in controlling electromagnetic waves in free space,[16,17] recent research has been expanded, exploiting MSs to govern guided waves[18-20] and further facilitating their interaction with freely propagating waves.[21,22] Surface waves, which are waves confined specifically to a surface, exhibit a rapid decrease in intensity with increasing distance from the surface.[20] These surface waves can be precisely and adaptably manipulated by properly designing MSs.[23,24] Hence, the utilization of MSs proves to be effective in transmitting energy and signals over long distances, particularly in wireless communications and wireless power transfer (WPT) systems. In the context of a typical WPT system, MSs demonstrated significant utility across various systems.[25-29] These systems are particularly beneficial in enhancing power transfer efficiency and extending the distance over which power can be transferred.

In particular, a limitation of conventional wireless power transfer based on free-space waves arises from the transmission range for high-frequency bands, where wireless signals spread over the surface of a sphere ($4\pi r^2$ with radius $r$) in free space and exhibit a transmission power inversely proportional to the squared propagation distance.[30] In particular, according to the Friis transmission equation,[31] high-frequency signals decay severely (i.e., proportional to the squared frequency). However, an MS can redirect the surface wave energy such that the energy propagates along the circumference of a circle ($2\pi r$). Thus, the transmitted energy density of an incident free-space wave ($T$) follows a relationship involving the input power ($P$) divided by the sphere area, expressed as $T \propto P/4\pi r^2$, whereas when the energy is transferred within a circle via a surface wave using an MS, the power decreases only as $T \propto P/2\pi r$. Therefore, we can reasonably expect an improvement in propagation efficiency when transferring power over longer distances.

Here, we propose an efficient surface wave propagation method based on a tape-based MS and demonstrate the performance numerically and experimentally. The proposed MS enables the propagation of electric field energy over an extended distance, as demonstrated through numerical simulations using a customized horn antenna and an integrated eigenmode solution within the ANSYS numerical simulator (Electronics Desktop 2022 R2).[32] In addition, theoretical fitting approaches and experimental validations are examined. The presence of patches made of conductive silver ink and embedded in the adhesive tape substrate alters the near-field and far-field interactions between the surface waves of the designed MS and the plane waves of free space, enhancing the coupling effect and resulting in an increase in the wave propagation distance. Moreover, our tape-based MSs are readily fabricated and applied to various environments thanks to the flexible adhesive tape layer. These tape-based MS devices form a robust foundation for future research on innovative wireless communication, signal processing, and related fields.

To provide a cost-effective solution, we use an adhesive tape substrate to advance large-area fabrication techniques via screen printing.[33] This kind of substrate offers enhanced flexibility, high precision, and excellent uniformity.[34-36] Furthermore, the inclusion of adhesive layers has become essential in numerous applications due to their role in accommodating manufacturing tolerances. As shown in Fig. 1(a), the structure consists of a periodic patch array made of conductive silver ink and patterned on an acrylic adhesive tape substrate. In the simulation, the geometric parameters of the proposed adhesive-tape-based MS include the periodicity $P$ (2, 3, or 4 mm), length of the silver square patch $L$ (=$P$-1 mm), and thickness of the adhesive substrate $t_s$ (0.5 mm).

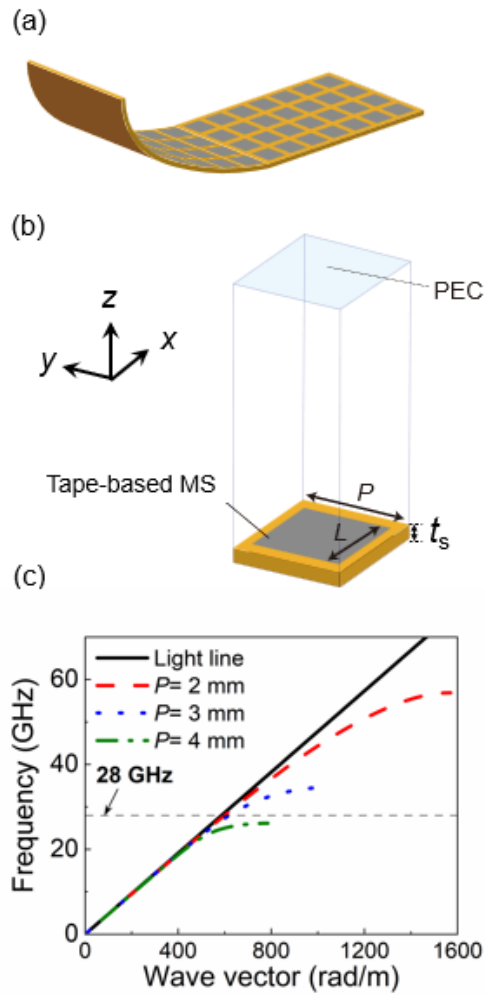

FIG. 1. Conceptual representation of a proposed tape-based MS and its fundamental characteristics. (a) Sketch of the flexible tape-based MS. (b) Detailed unit cell structure with a silver patch on the adhesive tape substrate. (c) Dispersion curves of different models of the proposed structure. The horizontal dashed gray line represents 28 GHz.

Fig. 1(b) shows the simulation model for eigenmode simulations. To investigate wave propagation in the *x*-direction, we use periodic boundary conditions for the *y-z* plane and *x-z* plane. The top and bottom boundary conditions are perfect electric conductor (PEC) walls. The dispersion relationship of the surface wave on the proposed MS is computed for various periodicities, as depicted in Fig. 1(c). The solid black line represents the vacuum light line. Different periodicities are illustrated through the red dashed line (*P*=2 mm), blue dotted line (*P*=3 mm), and green dashed-dotted line (*P*=4 mm). These curves gradually deviate from the light line and converge toward a resonant frequency, which indicates that the MS becomes a high-impedance surface. Therefore, while the surface wave propagation guided by the MS becomes notably limited near the resonant frequency, the fundamental mode of the surface wave is maintained below the resonant frequency.

To conduct a detailed analysis of surface wave propagation, we employ numerical simulations of the proposed MS in open space using radiation boundaries and a customized horn antenna.[37] This approach ensures that the surface wave attenuation is accurately extracted.[13] Fig. 2(a) illustrates the setup designed to observe surface wave propagation. The magnitude of the electric field versus the propagation distance for various MS samples is depicted in Fig. 2(b). In this analysis, electromagnetic wave propagation in free space can be categorized into two distinct regions according to the distance from the antenna aperture: the near-field and far-field regions. These regions determine the primary propagation domains of plane waves and surface waves, respectively. Specifically, the *E*-field intensity has high values at the antenna aperture (i.e., the propagation distance is equal to zero) and progressively decreases as the propagation distance increases, corresponding to the transitional region between plane waves and surface waves. For instance, the area within the range between 0 and 150 mm is mainly influenced by coexisting plane waves and surface waves, while the surface wave propagation becomes dominant from 150 to 300 mm. Specifically, the magnitude of the *E*-field is well maintained as the propagation distance is extended when *P* = 2 mm (red dashed line). However, when *P* = 3 mm (blue dotted line), a strong scattering of surface waves begins to occur due to the resonant frequency near 28 GHz. This resonant mechanism is more clearly observed when *P* = 4 mm (green dashed-dotted line), which strongly prevents the surface wave propagation, compared to other results including the result of the free-space wave propagation (black solid line).

The investigations reveal a decay of surface waves along the *x*-direction one-dimensionally. The corresponding field distribution over a two-dimensional surface is observed with various structures using *P* values of 2 mm, 3 mm, and 4 mm in Figs. 2(c)-(e), respectively, and is compared with the result without the MS array in Fig. 2(f).

These figures also support that the decay is notably associated with the excitation of surface plasmon resonances.[38] In Fig. 2(c), the surface wave is effectively guided and maintained with a large field intensity throughout the propagation distance. Similarly, the *E*-field energy is well-maintained with $P = 3$ mm, as shown in Fig. 2(d), but only within a short distance (150 mm) before experiencing rapid decline. Fig. 2(e) specifically demonstrates the absence of wave propagation in the *x*-direction with $P = 4$ mm due to the large surface impedance of the MS (see Fig. 1(c)). Likewise, Fig. 2(f) depicts limited propagation, but this poor performance is because the tape-based MS is not present, which implies that the wave propagation occurs mostly as a free-space wave rather than a surface wave.

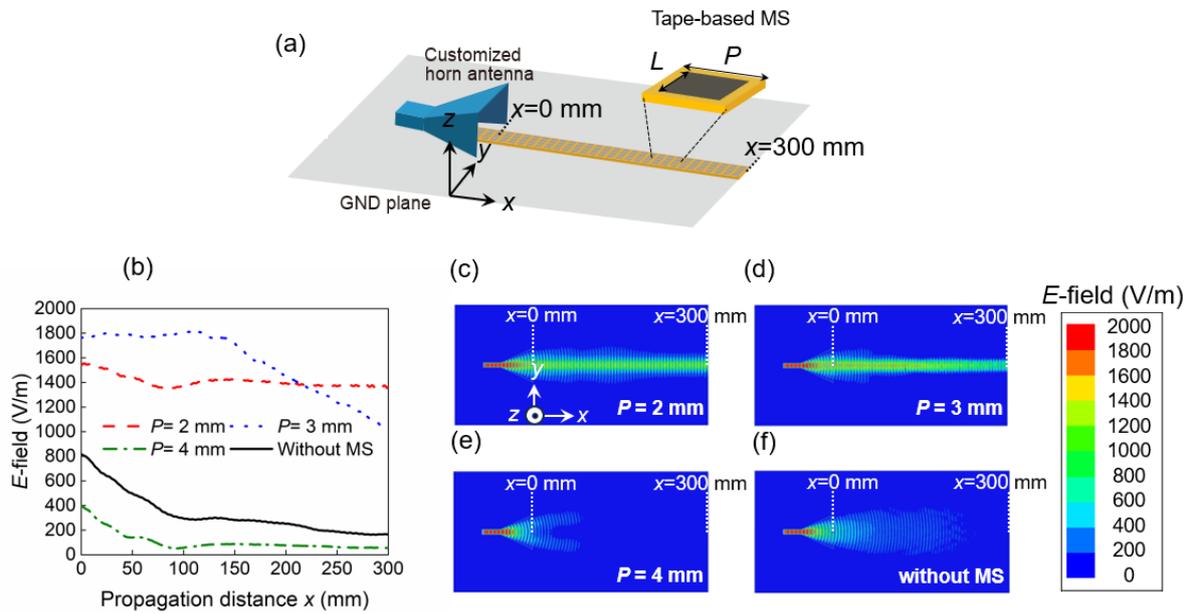

FIG. 2. Numerical validation of the tape-based MS. (a) Simulation model employing the customized horn antenna. (b) *E*-field calculated at 28 GHz for the proposed MS with various dimensions. (c)-(f) *E*-field distributions at 28 GHz with the MS using (c) $P = 2$ mm, (d) $P = 3$ mm, and (e) $P = 4$ mm and (f) without the MS. The *E*-field is calculated at 3 mm above the GND plane.

According to Figs. 2(b) and 2(c), the presence of the tape-based MS using $P = 2$ mm significantly extends the propagation distance of the surface wave. However, when *P* increases further, the propagation energy decreases due to the appearance of a large surface impedance (see Fig. 1(b)). Additional electric field profiles are depicted in Figs. S1-S3 of the **supplementary material** to confirm the presence of surface waves supported by the interface of the adhesive-tape-based MS structure across various frequencies in the high-frequency range.

For this experimental validation, we utilize a measurement system consisting of a customized horn antenna connected to an arbitrary waveform generator (AWG) (Keysight Technologies, M8195A) via coaxial cables, with the probe connected to a power sensor (PS) (Anritsu, MA2473D) and a power meter (PM) (Anritsu, ML2496A), as shown in Fig. 3(a). Additionally, an absorbing form is placed in front of the probe to reduce its unwanted scattering (see the blue part in Fig. 3(b)). Note that in this measurement system, electromagnetic field distributions are measured and estimated through the power value shown in the PM, which is converted to voltage to be readily compared with the *E*-field results shown in Fig. 2. The entire measurement process is conducted through our scanning system controlled by a LabVIEW program. Here, *E*-field distributions are measured in 2-mm steps over a sufficiently large area of 300 mm × 200 mm. The frequency of the incident wave is the same as the one adopted in the simulations of Fig. 2 (i.e., 28 GHz). Based on the simulation results of Fig. 2, we investigate the experimental performance of the proposed MS using *P*=2 mm, which is shown in Fig. 3(c). Additional measurement samples are fabricated and measured in Fig. S4 and Fig. S5 of the **supplementary material**.

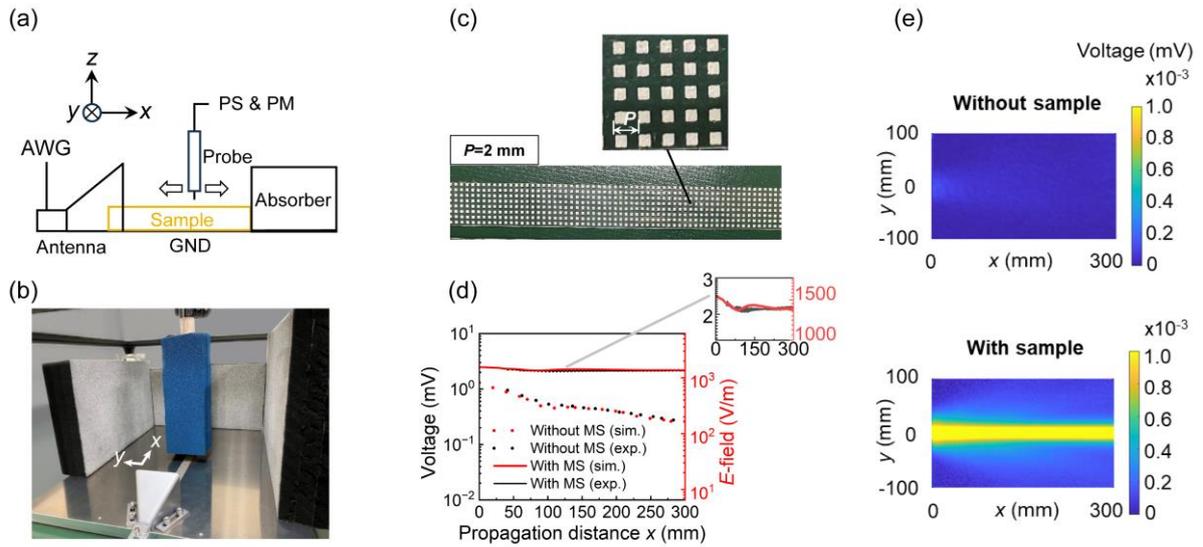

FIG. 3. Experimental validation of the tape-based MS. (a) Measurement setup and (b) photograph of our measurement system. (c) Sample considered in the measurement. Additional samples and corresponding results are presented in the supplementary material. (d) Comparison results between numerically calculated *E*-field (red lines) and experimentally obtained equivalent voltage (black lines) results with the MS (solid lines) and without the MS (dotted lines). (e) Equivalent voltage distribution with and without the MS.

The measurement result of the proposed MS is shown in Fig. 3(d) including the corresponding simulation result (i.e., *P*=2 mm of Fig. 2(b)). According to these results, a close agreement is obtained between the numerical and experimental results in both scenarios with and without the MS. In the presence of the MS, there is a gradual decline in power as the propagation distance increases, as shown in the enlarged graph of Fig. 3(d). Fig. 3(e) displays surface plots representing voltage values both with and without the MS. These results also support that the surface wave is strongly guided by our MS.

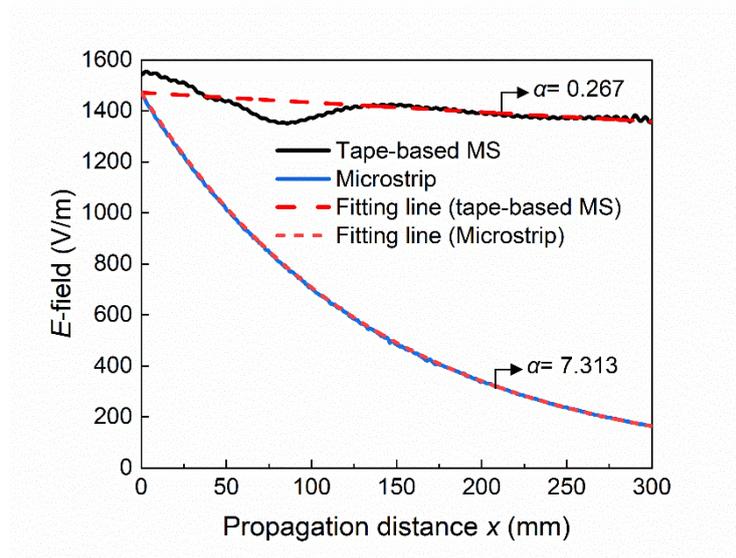

FIG. 4. Comparison of the performance of the tape-based MS and microstrip structure through the $E$-field energy against the propagation distance. The simulation results of the tape-based MS and the microstrip structure are indicated by the black solid line and blue solid line, respectively, while theoretical fitting results are shown by the red dashed lines for the taped-based MS and the dot red line for the microstrip.

In Fig. 4, a comparison is presented between the obtained results and fitting curves based on an exponential function:

$$E = A\exp(-\alpha x), \qquad (1)$$

where phase change is omitted for the sake of simplicity. Additionally, $A$ represents the $E$-field strength at $x=0$, and $\alpha$ denotes the attenuation constant. In Fig. 4, the attenuation value $\alpha$ serves as an indicator of the propagation efficiency of surface waves. Utilizing Eq. (1), we derive a small $\alpha$ value of 0.267 at 28 GHz for the tape-based MS, indicating well-maintained $E$-field energy against the propagation distance. Additionally, this result is compared with the performance of a microstrip—a conventional structure commonly used for guiding waves. This microstrip model is composed of the same materials as the ones adopted in the tape-based MS, as shown in Fig. S6 of the **supplemental material**. Interestingly, compared with the tape-based MS, a large $\alpha$ value of 7.313 is observed at the same frequency of 28 GHz for the microstrip structure. The significant difference between these attenuation constants indicates that the proposed MS is more efficient in transmitting the energy of the incoming signal with limited energy loss, compared to the microstrip line. The rationale for the improved performance of the MS is presumably because the MS does not fully confine the incident wave within the substrate, while the microstrip model has most of the energy within the substrate that has a dielectric loss to dissipate the incoming energy.

Our tape-based MS was numerically and experimentally demonstrated to show superior performance to transmit high-frequency signals as surface waves, compared to conventional approaches such as wireless propagation without the MS (Figs. 2(b) and 3(d)) and wired propagation with a microstrip (Fig. 4). However, the performance of the proposed MS can be further enhanced by introducing additional design concepts. For instance, surface waves can be strongly guided by changing the isotropic surface impedance to an anisotropic surface impedance.[39-43] The use of anisotropic MSs prevents surface waves from spreading over a two-dimensional surface and efficiently confines the surface wave energy along a one-dimensional direction only. In addition, transmitted signals can be preferentially selected by introducing lumped circuit components. Recent advancements in circuit-based MSs have introduced an additional level of control over signals at the same frequency with respect to the incident power and the waveform.[44-48] For instance, by using nonlinear circuit components within MSs, high-power destructive surface waves can be strongly absorbed to prevent sensitive wireless communication devices, while small signals are transmitted for communications at the same frequency.[44,46] Moreover, the same-frequency signals may be distinguished by sensing their pulse width, which is already leveraged in diverse fields, including antenna design,[49-51] electromagnetic interference,[52,53] wireless communications,[54-56] and signal processing.[56,57]

To summarize, we have reported MSs to efficiently transmit high-frequency signals as surface waves, as opposed to conventional wireless propagation and microstrip approaches. In particular, our MSs were designed with conductive silver inks patterned on an adhesive tape substrate for large-area production, flexibility, high precision, and excellent uniformity. Our MSs were numerically and experimentally evaluated to clarify the surface wave characteristics by using our comprehensive approach, which encompassed numerical simulations, theoretical fitting methods, and practical experiments within the 28-GHz band in free space. We used a customized horn antenna configuration for both numerical simulations and experiments, together with an automated scanning system for the measurements. Numerical and experimental results showed that the efficacy of the tape-based MS for wave propagation outperforms the results of conventional approaches such as wireless propagation without the MS and wired propagation with a microstrip. This study establishes a robust foundation for future research, offering insights into potential innovative applications in power-efficient next-generation wireless communications, wireless power transfer, signal processing, and beyond.

## SUPPLEMENTARY MATERIAL

See the supplementary material for the electric field profiles across various frequencies, as well as additional measurement results. Moreover, the configuration of the microstrip structure is provided.

## ACKNOWLEDGMENTS

This study was supported by the National Institute of Information and Communication Technology (NICT) of Japan under commissioned research No. JPJ012368C06201. SP acknowledged the support from Engineering and Physical Science Research Council Funding under EP/V048937/1.

## DATA AVAILABILITY

The data that support the findings of this study are available from the corresponding author upon reasonable request.

Supplementary Material for

# Experimental and Numerical Validation of Tape-Based Metasurfaces in Guiding High-Frequency Surface Waves for Efficient Power Transfer

K. Suzuki[1], P. T. Dang[1], H. Homma[1], A. A. Fathnan[1], Y. Ashikaga[2], Y. Tsuchiya[2], S. Phang[3], and H. Wakatsuchi[1]

[1]*Department of Engineering, Nagoya Institute of Technology, Gokiso-cho, Showa, Nagoya, Aichi, 466-8555, Japan*

[2]*Teraoka Seisakusyo Co., Ltd., 1-4-22 Hiromachi, Shinagawa-ward, Tokyo, 140-8711, Japan*

[3]*George Green Institute for Electromagnetics Research, Faculty of Engineering, University of Nottingham, University Park, Nottingham NG7 2RD, UK*

We numerically showed the *E*-field profiles of three models using $P$=2, 3, and 4 mm at 28 GHz, as depicted in Figs. 2(c)-(f). Figs. S1, S2, and S3 show simulation results of the *E*-field profiles corresponding to 22 GHz, 25 GHz, and 31 GHz, respectively.

At 22 GHz, both models using $P$=2 and 3 mm maintained the power of a surface wave over the simulated distance, as depicted in Fig. S1(a) and Fig. S1(b). However, with an increase in periodicity (see Fig. S1(c)), a resonant mechanism appears and markedly reduces the propagation distance. Fig. S1(d) illustrates limited propagation in the case without MS, which is attributed to the absence of surface waves.

A similar trend is observed at 25 GHz, as shown in Fig. S2. Both models using $P$=2 and 3 mm maintain the power level of the surface wave, as depicted in Fig. S2(a) and Fig. S2(b). Nevertheless, when the periodicity increases to $P$=4 mm (see Fig. S2(c)), the surface wave propagation is severely prevented.

At a higher frequency of 31 GHz (Fig. S3), propagation decay is evident in the model using $P$=3 mm due to pronounced surface wave attenuation. In contrast, the model using $P$=2 mm maintains consistent performance across the frequencies. This is because the model using $P$=2 mm still maintains a low surface impedance value, while the other models have high surface impedance values (see Fig. 1(c)). Therefore, a proper MS design is required to achieve excellent energy transfer efficiency for different operating frequencies.

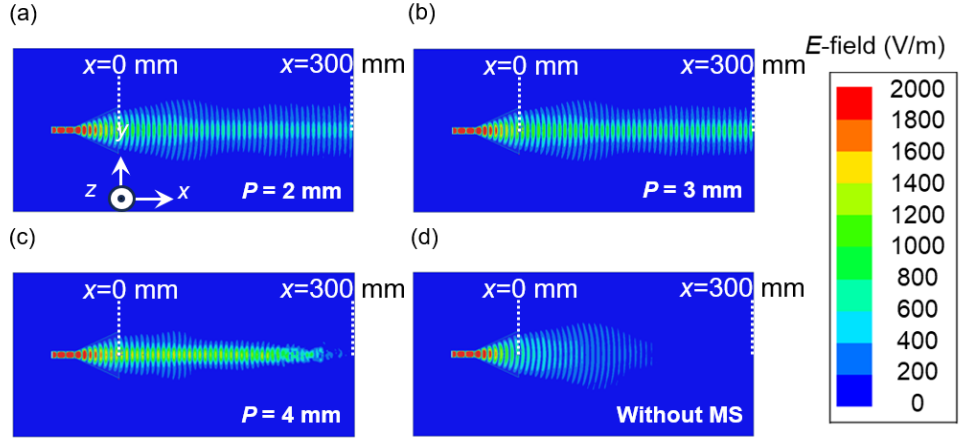

FIG. S1. Field distributions of various simulation models at 22 GHz. (a) $P$=2 mm, (b) $P$=3 mm, and (c) $P$=4 mm. (d) Result without MS.

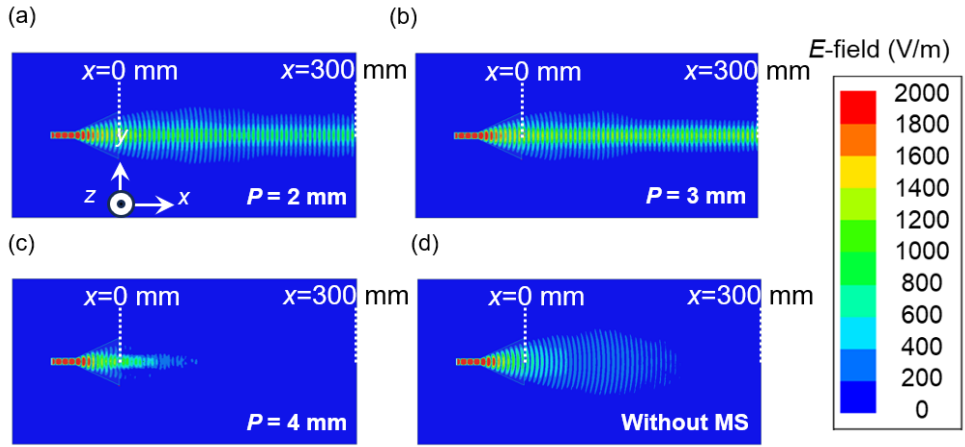

FIG. S2. Field distributions of various simulation models at 25 GHz. (a) $P$=2 mm, (b) $P$=3 mm, and (c) $P$=4 mm. (d) Result without MS.

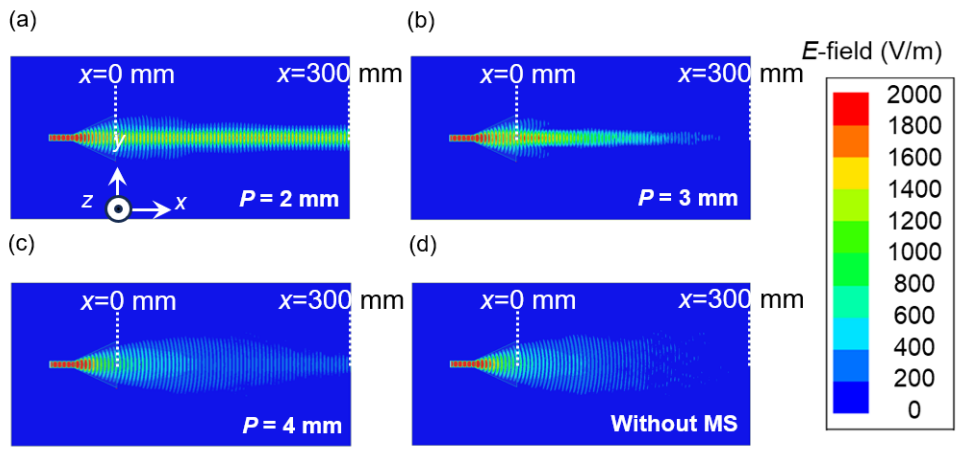

FIG. S3. Field distributions of various simulation models at 31 GHz. (a) $P$=2 mm, (b) $P$=3 mm, and (c) $P$=4 mm. (d) Result without MS.

We presented measurement results for the simulation model using $P$=2 mm at 28 GHz in the main manuscript, as displayed in Fig. 3(d). Additionally, we show measurement results for the simulation models using $P$=3 and 4 mm (corresponding to Figs. 2(d) and(e), respectively), as shown in Fig. S4. Fig. S4(a) shows measurement samples of the models using $P$=3 and 4 mm, respectively. Their measurement results are presented in Fig. S4(b) and Fig. S4(c). However, a partial deviation between the simulation and measurement results arises due to a sensitive resonant mechanism of the $P$=3 mm configuration, as the resonant frequency is located close to the measurement frequency of 28 GHz (refer to Fig. 1(c)). This resonant frequency shift changes strong surface wave scattering, as presented in Fig. 2(b) in the main manuscript, resulting in a decrease in the simulated surface wave power as the propagation distance increases. Similarly, when the periodicity increases to $P$= 4 mm, the strength of the sensitive resonance is influenced by the actual physical dimension, which appears as a further reduction in the transmitted voltage at 28 GHz. Consequently, the surface wave power cannot be maintained over the propagation distance, as shown in Fig. S4(c). These differences due to the resonant frequency shift can be explained by Fig. S5, where the measurement samples using $P$=2, 3, and 4 mm are tested at a lower frequency of 25 GHz. With $P$=2 mm and $P$=3 mm, the structural resonant frequencies of these configurations are sufficiently distant to avoid unwanted scattering. Therefore, the simulation and measurement results show good agreement, as depicted in Figs. S5(a) and S5(b). On the other hand, when $P$=4 mm, the frequency of the sensitive resonant mechanism is close to 25 GHz so that a clear deviation is found between the simulation and measurement results, as shown in Fig. S5(c). These results are compared to the dispersion diagram of Fig. 1(c), which shows that the result of $P$=4 mm starts deviating from the light line near 25 GHz. Despite such differences, still these additional measurement results support that the proposed tape-based MS enables efficient wave propagation by properly adjusting the physical dimensions.

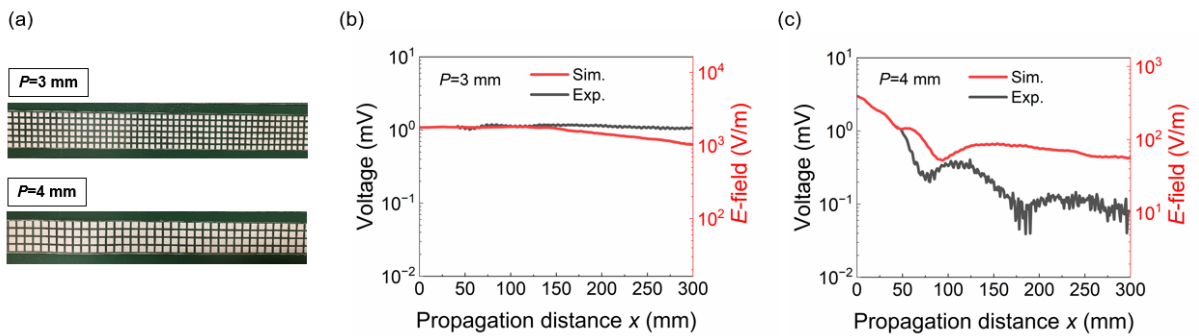

FIG. S4. Additional experimental results of the proposed tape-based MS using different periodicities at 28 GHz. (a) Additional measurement samples using $P$=3 and 4 mm. (b-c) Comparison between the simulation results (red) and the experimental results (black) for $P$=3 mm (b) and 4 mm (c).

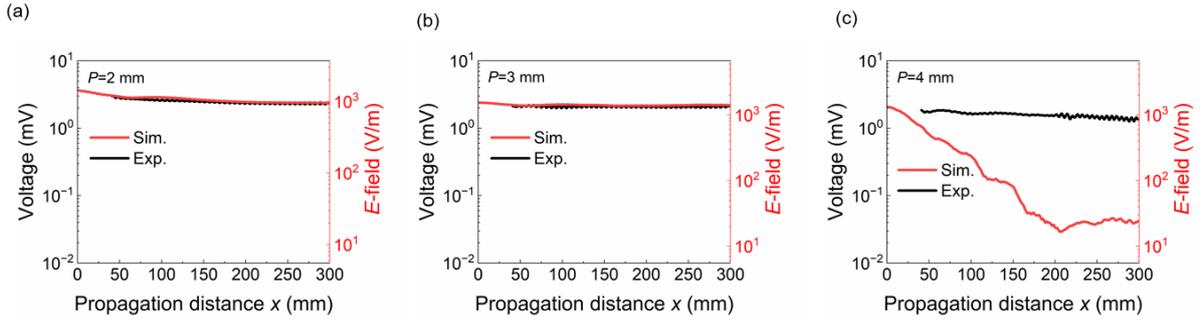

FIG. S5. Additional experimental results at 22 GHz and corresponding simulation results. (a) *P*=2 mm, (b) *P*=3 mm, and (c) *P*=4 mm.

In addition, we compared the attenuation constant of the tape-based MS with that of a microstrip model at 28 GHz to demonstrate efficient propagation exceeding one of conventional approaches, as shown in Fig. 4. The simulation model used for the microstrip is displayed in Fig. S6. The substrate thickness, width, and length are set to 0.5, 20, and 50 mm, respectively, and the silver line width is 1.6 mm. The dielectric properties of the substrate and the conductivity of the silver are all the same as the conditions applied to the proposed MS.

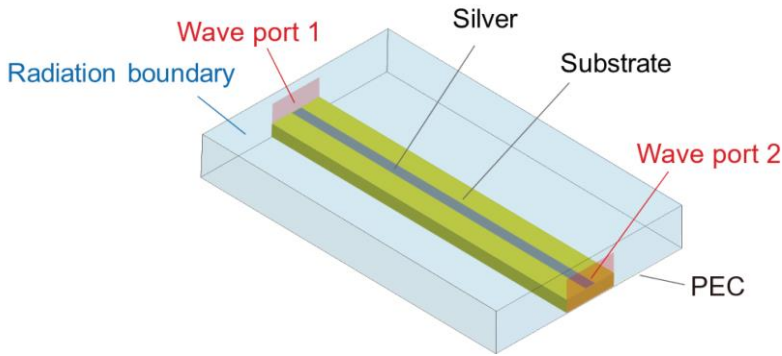

FIG. S6. Sketch of the microstrip simulation model used in Fig. 4.